\documentclass[twocolumn]{aastex62}
\usepackage{amsmath,amstext}
\usepackage[T1]{fontenc}
\usepackage{apjfonts} 
\usepackage[figure,figure*]{hypcap}
\usepackage{commath}

   \newcommand{\beam}{$\theta_{\mbox{\scriptsize maj}}\times\theta_{\mbox{\scriptsize min}}$}

   \newcommand{\amax}{ $a_{\mbox{\scriptsize max}}$ }

\shorttitle{Millimeter continuum spectral index in TW Hya}
\shortauthors{Liu, H.-B.}

\begin{document}

\title{The anomalously low (sub)millimeter spectral indices of some protoplanetary disks may be explained by dust self-scattering}

\correspondingauthor{Hauyu Baobab Liu}
\email{hyliu@asiaa.sinica.edu.tw}

\author[0000-0003-2300-2626]{Hauyu Baobab Liu}
\affiliation{Academia Sinica Institute of Astronomy and Astrophysics, P.O. Box 23-141, Taipei 10617, Taiwan}

\begin{abstract}
Previous (sub)millimeter observations have found that the spectral indices of dust emission from some young stellar objects are lower than that of the black body emission in the Rayleigh-Jeans limit (i.e., 2.0).
In particular, the recent Atacama Large Millimeter Array observations have spatially resolved that the innermost regions of the protoplanetary disks  TW\,Hya and HD\,163296 present anomalously low (i.e., $<$2.0) millimeter spectral indices.
In some previous works, such anomalously low millimeter spectral indices were considered unphysical and were attributed to measurement errors.
The present work clarifies that if the albedo is high and is increasing with frequency, it is possible to reproduce such anomalously low spectral indices when the emission source is optically thick.
In addition, to yield lower than 2.0 spectral index at (sub)millimeter bands, the required dust maximum grain size \amax is on the order of 10-100 \micron, which is well-consistent with the previously derived \amax values based on multi-wavelength dust polarimetric observations.
In light of this, measuring Stokes I spectral index may also serve as an auxiliary approach for assessing whether the observed dust polarization is mainly due to dust scattering or is due to the aligned dust grains.
\end{abstract}

\keywords{stars: individual (TW\,Hya) --- protoplanetary disks}

\section{Introduction}\label{sec:introduction}
To approximate interstellar dust emission, the so called modified black body formulation has been widely applied (for a review see \citealt{Hildebrand1983}) 
\[ S_{\nu}=\Omega B_{\nu}(T_{\mbox{\scriptsize dust}})(1-e^{-\tau_{\nu}}),\] 
where $S_{\nu}$ is the observed flux at frequency $\nu$, $\Omega$ is the solid angle of the emission region, $B_{\nu}(T_{\mbox{\scriptsize dust}})$ is the Planck function at dust temperature $T_{\mbox{\scriptsize dust}}$ and frequency $\nu$, and $\tau_{\nu}$ is the optical depth of dust.
The dust optical depth $\tau_{\nu}$ is further expressed as the product of the dust mass absorption opacity ($\kappa_{\nu}^{\mbox{\scriptsize abs}}$) and dust mass surface density ($\Sigma$).
Since dust grains cannot emit or absorb efficiently at wavelengths which are much longer than their size, at (sub)millimeter bands, $\kappa_{\nu}^{\mbox{\scriptsize abs}}$ is proportional to $(\nu)^{\beta}$, where $\beta$ is known as the dust opacity spectral index.
The value of $\beta$ is $\sim$2 in the diffuse interstellar medium (ISM) around the solar neighborhood. 
With the presence of dust that has grown larger, the value of $\beta$ can become as low as 0.0.
In the Rayleigh-Jeans limit, the observed (sub)millimeter spectral index ($\alpha$) is related to $\beta$ by $\alpha=\beta+2$.

Some previous (sub)millimeter observations of protoplanetary disks have reported $\alpha\sim$2.5.
By assuming that the dust scattering opacity ($\kappa_{\nu}^{\mbox{\scriptsize sca}}$) is negligible, they argued that $\beta\sim$0.5 and suggested that millimeter-sized dust grains may already present in those disks \citep[e.g.,][and references therein]{Beckwith1991ApJ,Carrasco2016ApJ}.
However, lately some observations reported anomalously low (sub)millimeter spectral indices ($\alpha<$2.0), which are inconsistent with the aforementioned formulation of interstellar dust emission (e.g., Class 0/I objects: \citealt{Jorgensen2007ApJ,Miotello2014A&A,Li2017ApJ,Liu2018A&A1,AgurtoGangas2019}; protoplanetary disks: \citealt{Tsukagoshi2016,Liu2017A&A,Huang2018,Dent2019}).
Due to the significant numbers of such reports, some of which were carried out by teams which possess authority on the technical ground \citep[e.g.,][]{Dent2019}, it is hard to attribute all of them to data calibration errors or imaging artifacts.
Another related paradox is that recent, multi-wavelength polarimetric observations of dust scattering \citep[for more details of this mechanism see][]{Kataoka2015ApJ, Yang2017MNRAS} mostly concluded that the maximum grain sizes \amax are  $\sim$50-150 $\mu$m, and are not yet fully reconciled with those earlier  suggestions of millimeter-sized grains based on analyzing spectral indices $\alpha$ \citep[see][]{Kataoka2016ApJa,Kataoka2016ApJb,Stephens2017ApJ,Bacciotti2018ApJ,Hull2018}.

Based on radiative transfer models, \citet{Li2017ApJ} and \citet{Galvan2018ApJ} have argued that when dust grains are small ($\ll$1 mm), the anomalously low $\alpha$ values can be explained by the presence of  foreground obscured hot dust.
Otherwise, low values of $\alpha$ may be explained by a component of free-free emission \citep[e.g.][]{Liu2017A&A}.
\citet{Li2017ApJ} and \citet{Galvan2018ApJ} found that applying foreground obscured hot dust better explains the (sub)millimeter spectral energy distributions (SEDs) of some Class 0/I young stellar objects (YSOs) observed on 100-1000 AU scales.

Our present understanding, however, is that dust in Class\,II protoplanetary disks is predominantly heated by protostellar irradiation.
Therefore, dust around the disk surface is likely hotter than that at the disk midplane.
If this is indeed the case, then the explanation of foreground obscured hot dust cannot be applied to the Class\,II protoplanetary disks which are observed in face-on projection.
On the other hand, time monitoring observations \citep[e.g.,][]{Galvan2014,Liu2014ApJ} have shown that the free-free and/or synchrotron emission from Class II protoplanetary disks are rarely bright enough to be able to confuse the measurements of dust emission at (sub)millimeter bands.
In light of these, it is particularly puzzling that the low values of  $\alpha$ ($<$ 2.0) have been spatially resolved in the inner $\lesssim$10 AU radii of the approximately face-on, low-luminosity protoplanetary disk TW\,Hya (initially reported by \citealt{Tsukagoshi2016}, and was reproduced by \citealt{Huang2018} with independent measurements), and from HD\,163296 \citep{Dent2019}.

Based on simplified radiative transfer models, the work presented here argues that if we take scattering opacity into consideration, it is possible to reproduce the anomalously low $\alpha$ values at (sub)millimeter bands from an isothermal, high optical depth dust emission source with \amax$\sim$0.1\,mm.
The analysis will be compared specifically to Atacama Large Millimeter Array (ALMA) observations of the Class\,II protoplanetary disk, TW\,Hya \citep[$d\sim$60 pc;][]{Gaia2016A&A,Gaia2018A&A}.
Thanks to its approximately face-on projection \citep[for more information of this target source see][and references therein]{Qi2004ApJ, Andrews2016ApJ}, it may be sufficient to consider the analytic solution of radiative transfer equation for a thin slab, without requiring full three-dimensional Monte Carlo radiative transfer modeling.
Therefore, the analysis can be based on fewer free parameters and the results would remain robust and comprehensive.
In addition, there is less concern about the confusion of free-free emission thanks to the low bolometric luminosity and low protostellar mass of TW\,Hya.

The observational data used in this work are briefly introduced in Section \ref{sub:alma} while more details are given in Appendix \ref{appendix}.
The analysis of the SEDs is provided in Section \ref{sub:sed}.
Section \ref{sec:discussion} discusses the general implication of this work to other observational case studies, while our conclusion is nearly identical to the Abstract.



\begin{figure}
    \hspace{-1.5cm}
    \begin{tabular}{c}
         \vspace{-1cm}\includegraphics[width=10cm]{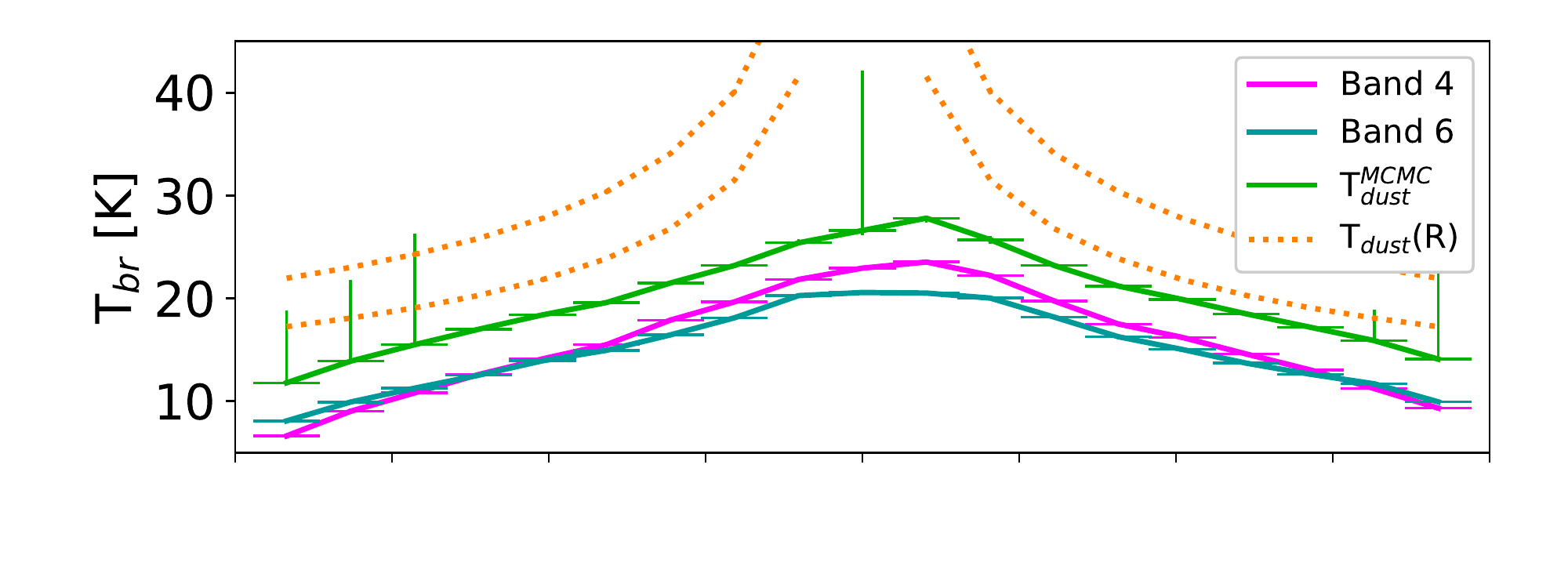} \\
         \vspace{-1cm}\includegraphics[width=10cm]{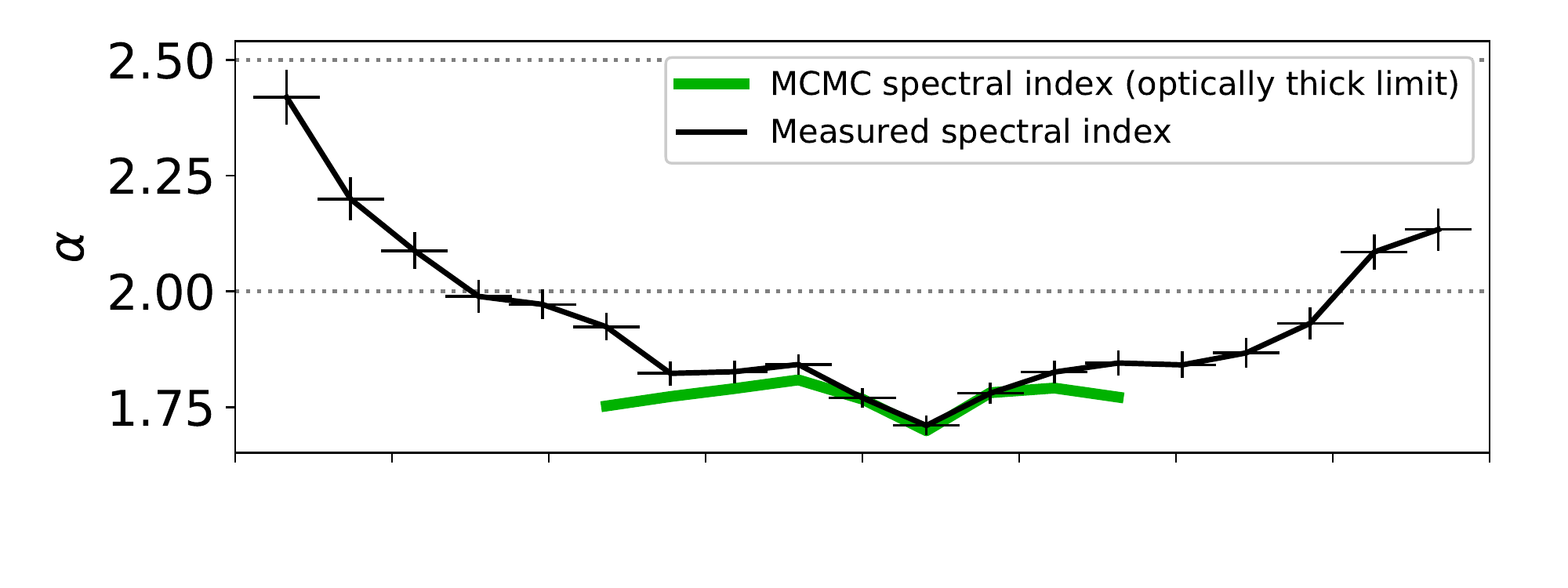} \\
         \includegraphics[width=10cm]{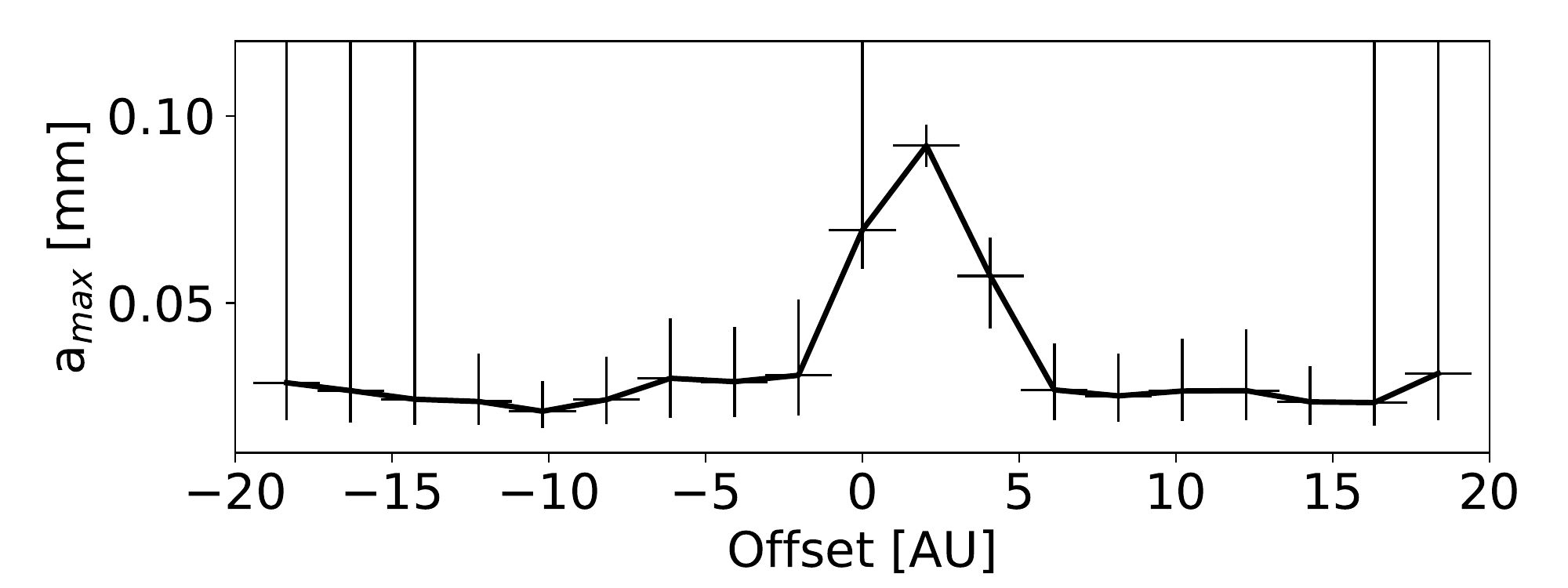} \\
    \end{tabular}
    \caption{Profiles of the dust brightness temperature ($T_{\mbox{\scriptsize br} }$) observed at Bands 4 and 6 ({\it top}), the Band 4-6 spectral indices derived from observations and MCMC fittings ($\alpha$; {\it middle}), and the dust temperature $T_{\mbox{\scriptsize dust}}^{\mbox{\scriptsize MCMC}}$ ({\it top}) and  maximum grain size (\amax; {\it bottom}) profiles derived from MCMC fittings, which were measured along the major axis of TW\,Hya (P.A.=155$^{\circ}$; positive offset is defined towards the southeast).
    The vertical error bars of the MCMC fitting results present the 25th and the 75th percentiles.
    Top panel also shows the power-law $T_{\mbox{\scriptsize dust}}(R)$=22 [K]$\times$(R/10 [AU])$^{-0.4}$ and 28 [K]$\times$(R/10 [AU])$^{-0.4}$ temperature models suggested by \citet{Andrews2012ApJ,Andrews2016ApJ}.
    }
    \label{fig:profiles}
\end{figure}

\section{Data Analysis}\label{sec:analysis}
\subsection{Millimeter spectral index from TW\,Hya}\label{sub:alma}
The ALMA Band 4 ($\sim$145 GHz) and Band 6 ($\sim$233 GHz) data taken from project 2015.A.00005.S (PI: Takashi Tsukagoshi), and the ALMA Band 6 data taken from project 2013.1.00114.S (PI: Karin \"Oberg)\footnote{Note that \citet{Tsukagoshi2016} referred to project 2012.1.00422.S instead of 2013.1.00114.S, which was likely a typo since 2012.1.00422.S did not carry out Band 6 observations.} were used for the present work.
These two bands are ideal for the present science purpose due to the sufficiently high dust optical depths, and and because both weavelengths can be approximated by the Rayleigh-Jeans limit.
That the SED analysis could become degenerate when mixing non-Reyleigh-Jeans and Rayleigh-Jeans components is a concern.
More details about how the data calibration was reproduced are given in Appendix \ref{appendix}.

Top and middle panels of Figure \ref{fig:profiles} present the dust brightness temperature $T_{\mbox{\scriptsize br}}$ and spectral index ($\alpha$) taken from a thin slice along the major axis (P.A.=155$^{\circ}$, see \citealt{Qi2004ApJ,Andrews2012ApJ}) of TW\,Hya.
The peak $T_{\mbox{\scriptsize br}}$ value detected in this work is lower than that in \citet{Tsukagoshi2016}, which is likely due to the poorer angular resolution adopted in this work.
In addition, this work did not perform azimuthal averaging to avoid smearing the weakly resolved azimuthal asymmetry at the innermost ring (see Figure 1 of \citealt{Tsukagoshi2016}; an also see \citealt{Roberge2005ApJ} for a related claim on large spatial scales).
Beyond these minor deviations, the results presented in Figure \ref{fig:profiles} largely agree with what was presented in \citet{Tsukagoshi2016}.
The anomalously low (i.e., $<$2.0) $\alpha$ values were reproduced in the inner $\sim$10 AU radii.

\begin{figure}
  \hspace{-1.2cm}
  \begin{tabular}{c}
    \includegraphics[width=9cm]{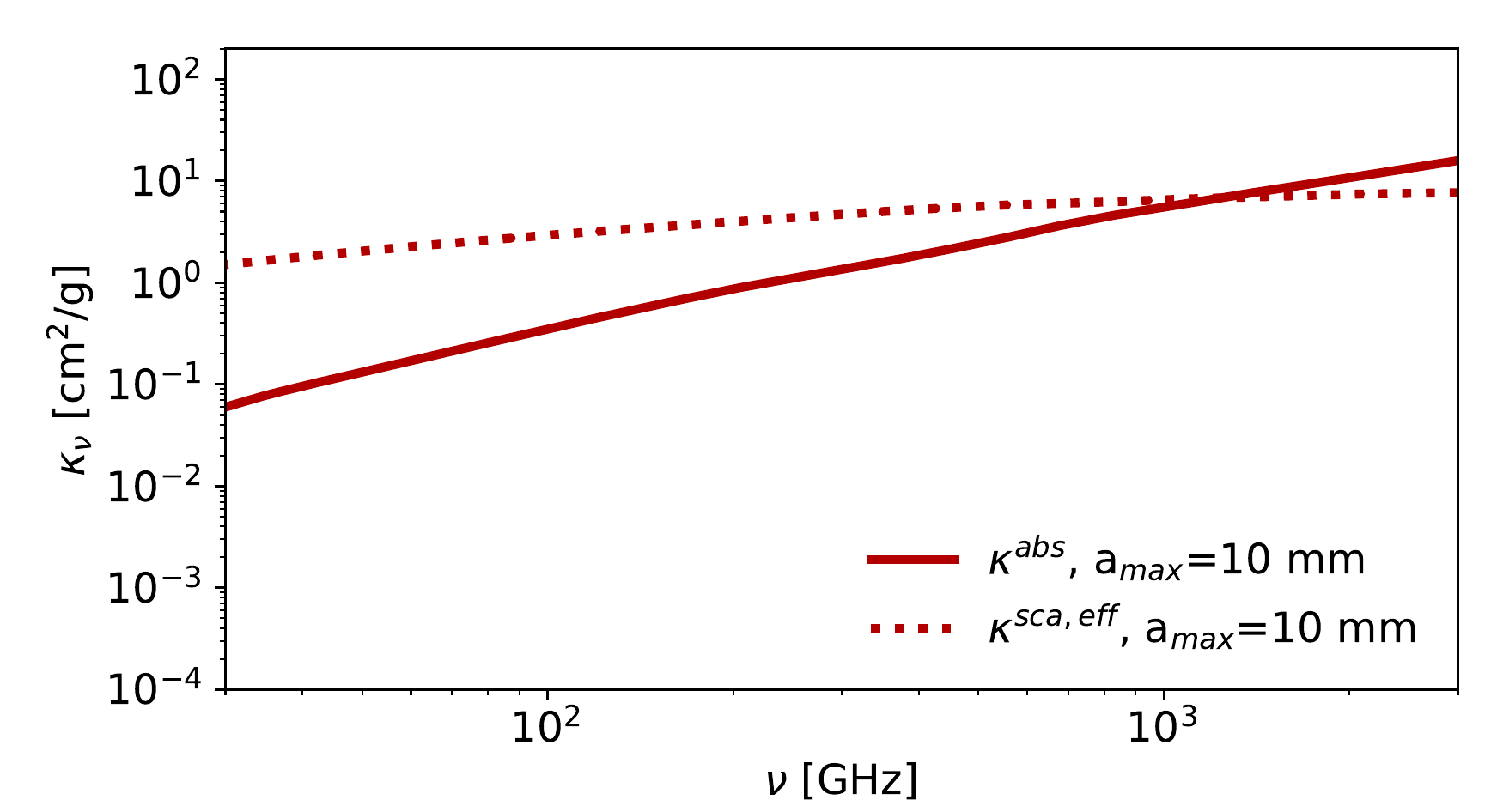}\vspace{-0.4cm} \\\vspace{-0.4cm}
    \includegraphics[width=9cm]{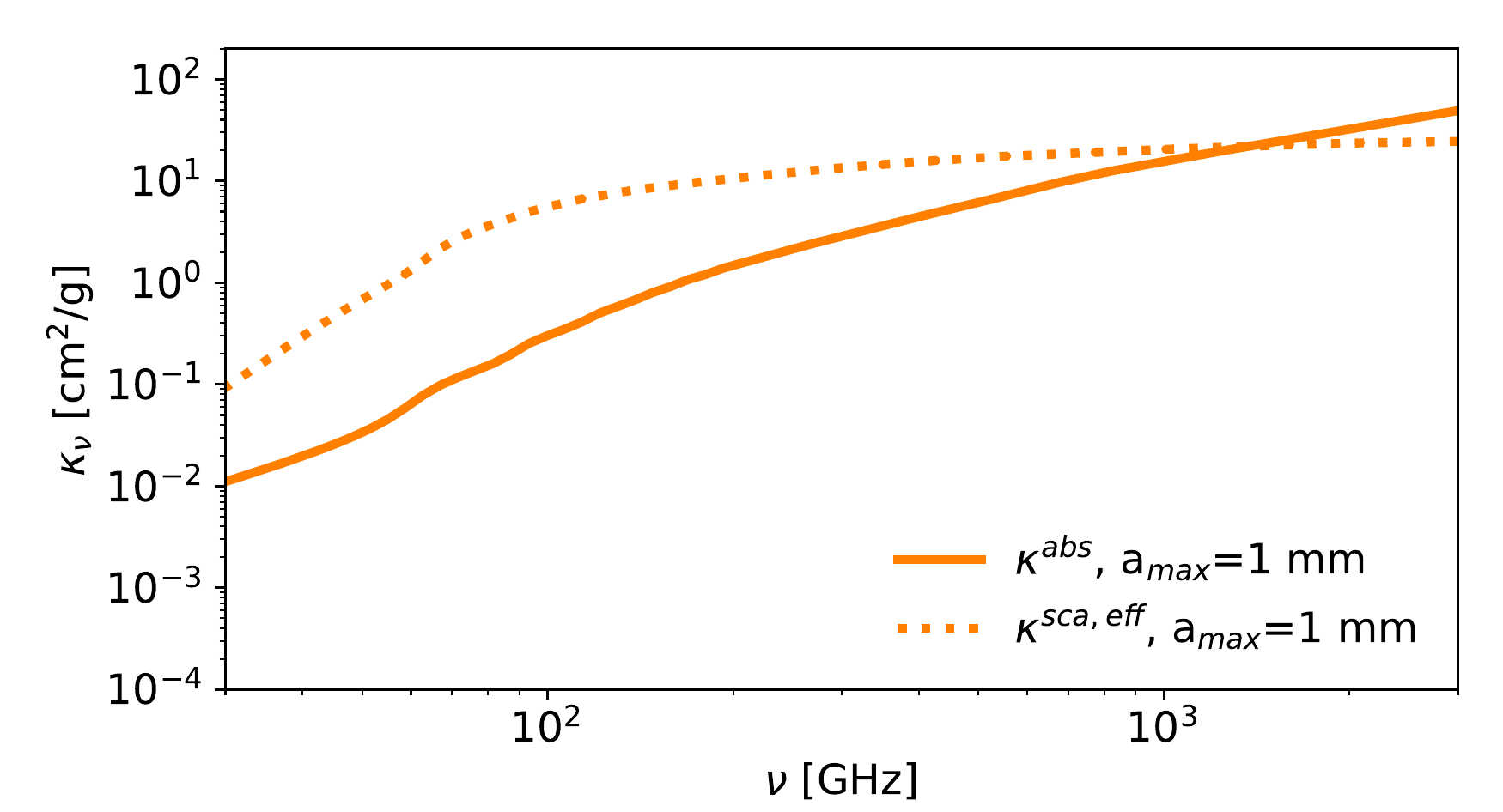} \\\vspace{-0.4cm}
    \includegraphics[width=9cm]{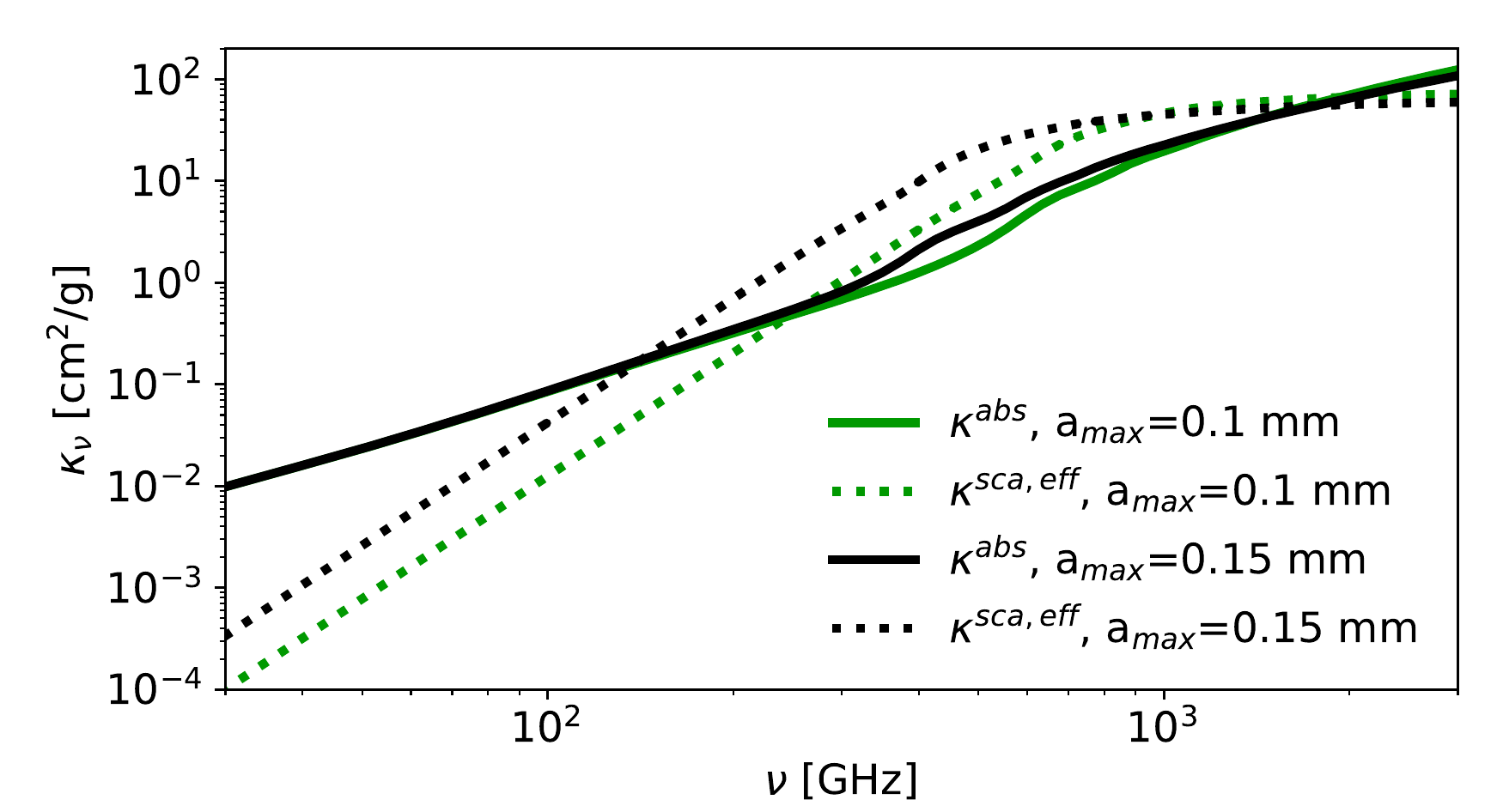} \\\vspace{-0.4cm}
    \includegraphics[width=9cm]{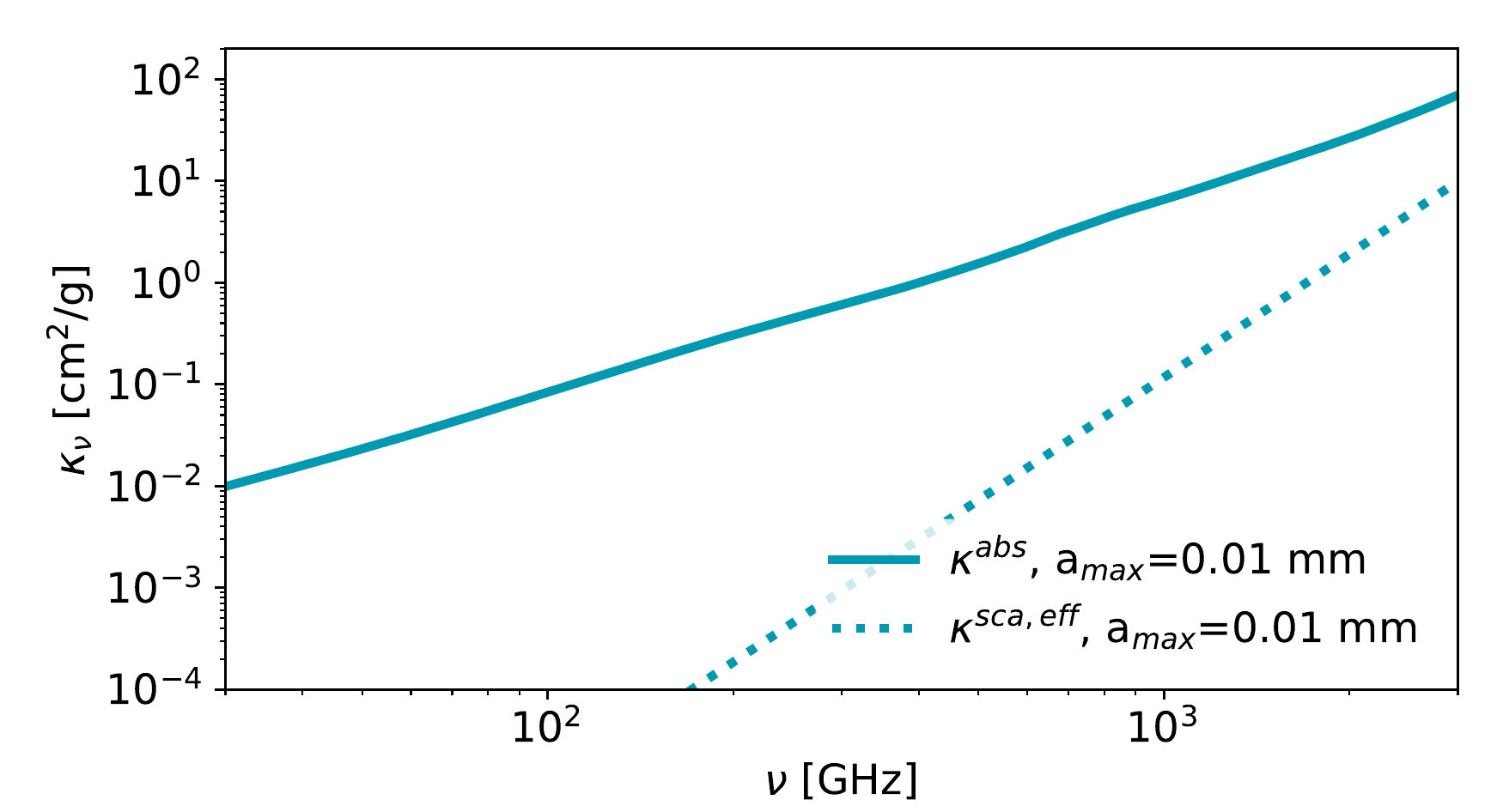} \\\vspace{0.1cm}
  \end{tabular}
  \caption{Absorption ($\kappa^{\mbox{\scriptsize abs}}$) and approximated scattering ($\kappa^{\mbox{\scriptsize sca,eff}}$) opacity of dust derived assuming the DSHARP optical constants \citep{Birnstiel2018} and a power-law size ($a$) distribution (i.e., $n(a)\propto a^{-q}$) in between the assumed minimum and maximum grain sizes $a_{\mbox{\scriptsize min}}$, \amax.
  This work adopted the minimum grain size $a_{\mbox{\scriptsize min}}=$ 10$^{-4}$ mm and the power-law index $q$=3.5. From top to bottom panels shows the cases with \amax=10, 1, 0.1, and 0.01 mm, respectively. The third panel also shows the case with \amax=0.15 mm to demonstrate how the variation rate of albedo may be sensitive to small changes in \amax.}
  \label{fig:opacity}
\end{figure}

\begin{figure}
    \vspace{-0.4cm}
    \hspace{-0.5cm}
    \begin{tabular}{ p{9cm} }
      \hspace{-0.3cm}\includegraphics[width=8.9cm]{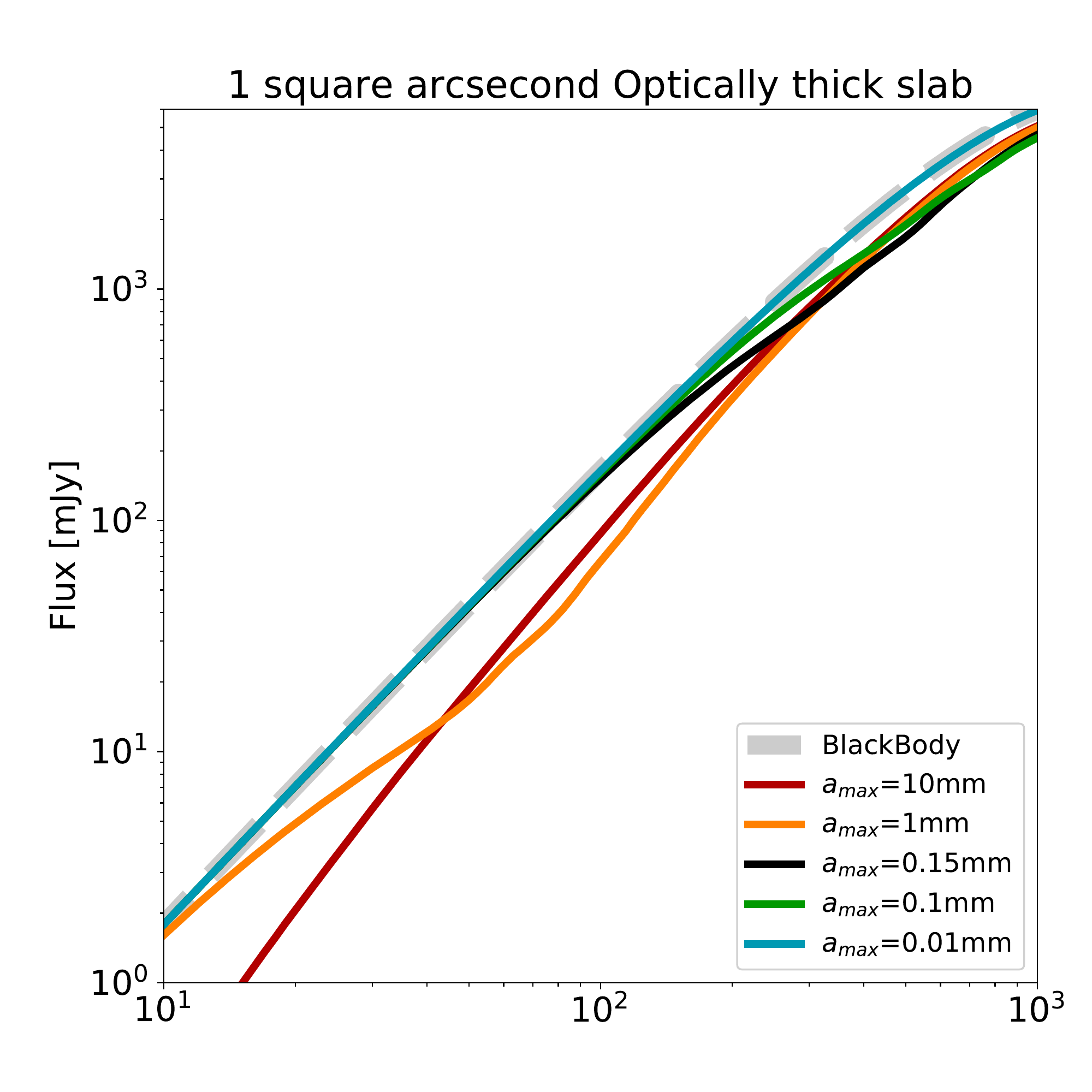} \\
      \vspace{-1.75cm}\includegraphics[width=9cm]{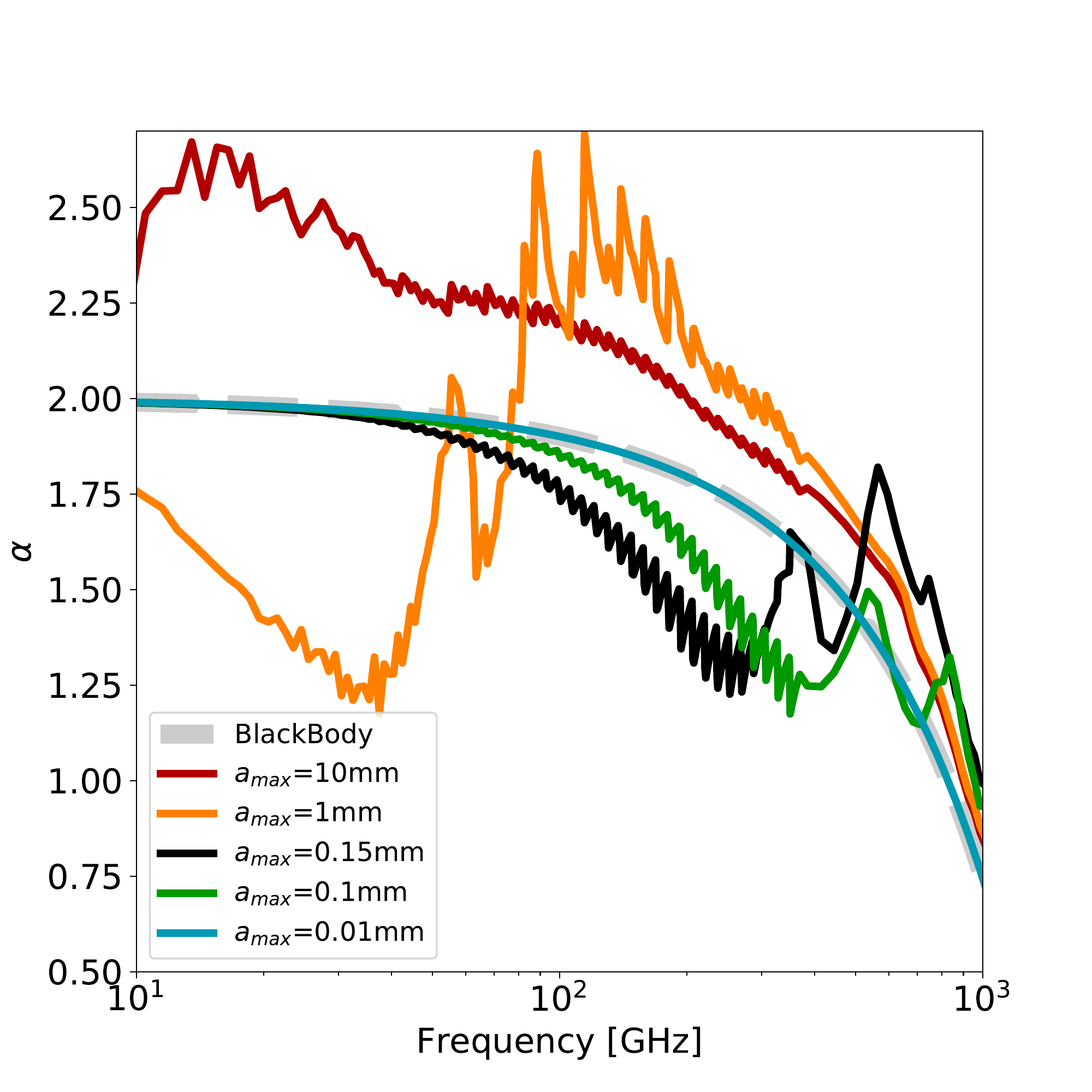} \\
    \end{tabular}
    \vspace{-0.3cm}
    \caption{Examples of the (sub)millimeter spectral energy distribution (top) and spectral index (bottom) evaluated for an optically thick ($\tau\gg1$) isothermal (25 K) dust slab of 1 square arcsecond angular scale. Gray dashed line shows the case of black body emission. Solid lines show the cases evaluated based on the assumption of the DSHARP opacities presented in Figure \ref{fig:opacity}.}
    \label{fig:sedexample}
\end{figure}

\subsection{Estimating maximum grain sizes based on fitting spectral index}\label{sub:sed}
This work adopted the default DSHARP dust optical constants published in \citet{Birnstiel2018}, which appears qualitatively similar to what was presented in the independent work of \citet{Kataoka2015ApJ}.
The ice-free dust opacity was not considered in this work since the detected dust brightness temperature is well below the typically assumed sublimation temperature for water ice (100-200 K).
The size averaged dust absorption ($\kappa_{\nu}^{\mbox{\scriptsize abs}}$) and effective scattering ($\kappa_{\nu}^{\mbox{\scriptsize sca,eff}}$) opacities were evaluated based on an assumption of spherical compact grains, a power-law grain size distribution with a power-law index $q=$3.5, a minimum grain size $a_{\mbox{\scriptsize min}}=$10$^{-4}$ mm and a maximum grain size \amax, the Mie theory and the Henyey-Greenstein scattering approximation.
Figure \ref{fig:opacity} shows the examples $\kappa_{\nu}^{\mbox{\scriptsize abs}}$ and $\kappa_{\nu}^{\mbox{\scriptsize sca,eff}}$ for \amax=0.01, 0.1, 0.15, 1.0, and 10 mm, respectively.
From this figure, we can see that dust scattering is negligible when \amax is $<$0.01 mm.
In addition, at long wavelength $\kappa_{\nu}^{\mbox{\scriptsize sca,eff}}$ has steeper slope than $\kappa_{\nu}^{\mbox{\scriptsize abs}}$.
When \amax is close to $\sim$0.1 mm, $\kappa_{\nu}^{\mbox{\scriptsize sca,eff}}$ becomes comparable with $\kappa_{\nu}^{\mbox{\scriptsize abs}}$, and the curve of $\kappa_{\nu}^{\mbox{\scriptsize sca,eff}}$ has an intersection with the curve of $\kappa_{\nu}^{\mbox{\scriptsize abs}}$ at the steep slope tail of $\kappa_{\nu}^{\mbox{\scriptsize sca,eff}}$ (e.g., at $\gtrsim$1 mm wavelengths).
Therefore, around this intersection, there is a range of wavelength where the albedo is not negligible and is rapidly decreasing with wavelength.
This feature moves to centimeter wavelengths when \amax$\gtrsim$1 mm.
As a consequence of a higher fraction of dust emission  scattering off at shorter wavelengths, at wavelengths close to the aforementioned $\kappa_{\nu}^{\mbox{\scriptsize abs}}$-$\kappa_{\nu}^{\mbox{\scriptsize sca,eff}}$ intersection the $T_{\mbox{\scriptsize br}}$ of an optically thick isothermal dust slab will increase with wavelength (c.f. Figure 9 of \citealt{Birnstiel2018}).
Figure \ref{fig:sedexample} shows examples of the (sub)millimeter SEDs for various values of \amax, evaluated based on the analytic radiative transfer solution for such optically thick, isothermal (25 K), geometrically thin dust slab in face-on projection, which was introduced in \citet{Birnstiel2018}. 
They are compared with the ordinary black body SED which has $\alpha$=2.0 in the Rayleigh-Jeans limit.
Here we can clearly see that when dust scattering is taken into account, and when \amax$\sim$0.1 mm, the anomalously low values of  $\alpha<$2.0 are reproduced at millimeter wavelengths, manifesting as a flatter SED than that of the ordinary black body.
The feature of anomalously low $\alpha$ shifts to centimeter wavelengths when \amax$\sim$1 mm.

\begin{figure}
    \hspace{-1cm}
    \includegraphics[width=10cm]{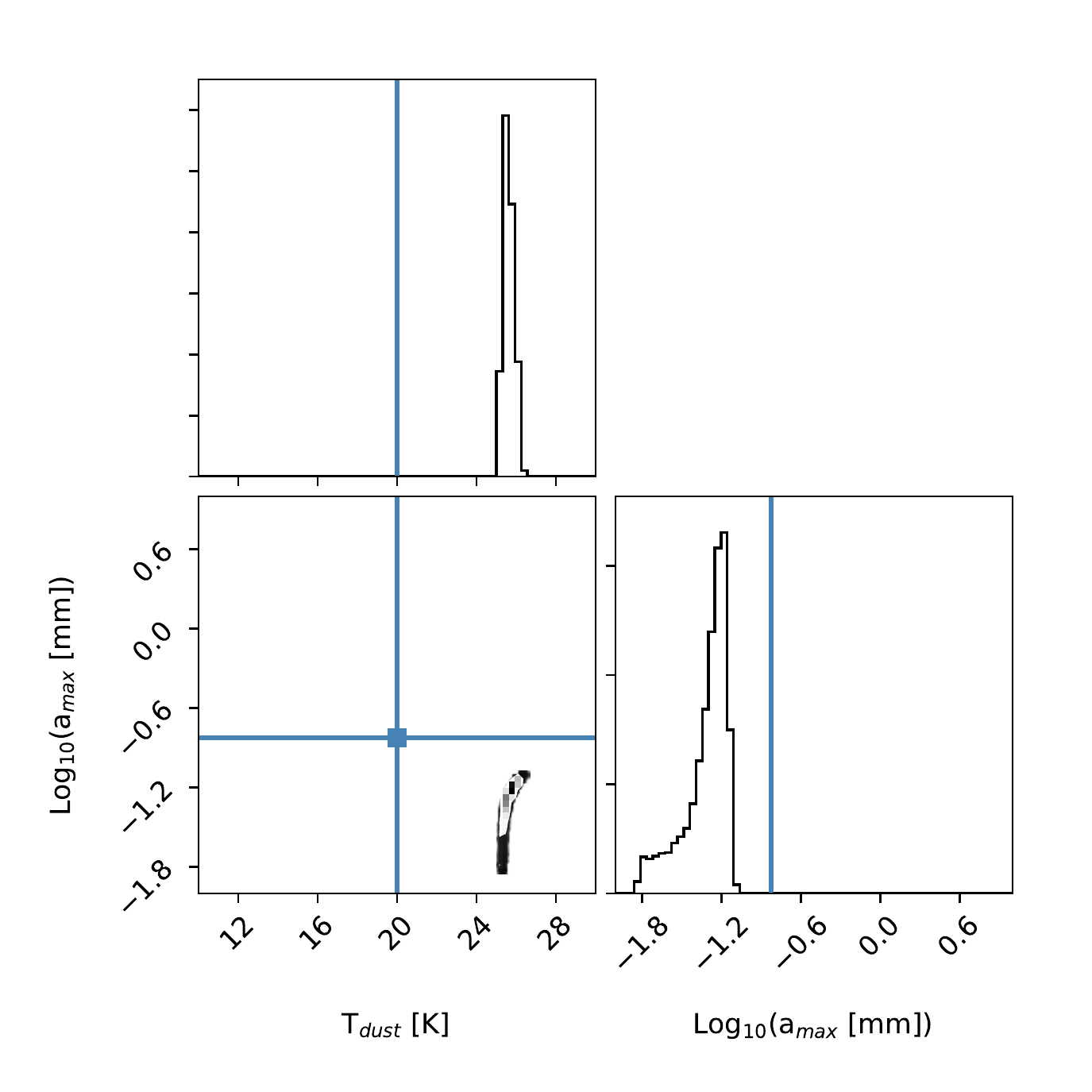}
    \vspace{-0.8cm}
    \caption{Corner plot for the results of MCMC fittings at 4.1 AU offset. Blue lines show the mean initial position of the MCMC walkers.}
    \label{fig:corner}
\end{figure}

 Markov chain Monte Carlo (MCMC) fittings to the $T_{\mbox{\scriptsize br}}$ profiles of TW\,Hya (Figure \ref{fig:profiles}) were carried out to examine what \amax values are indicated by the optically thick ($\tau\gg$1), isothermal and face-on thin dust slab model adopted here.
 In this case, MCMC is easier to implement than other fitting methods since every iteration of fittings needs to re-evaluate dust opacities based on the advanced \amax value.
 The MCMC fittings were initialized with 100 walkers at the mean initial positions of [$T_{\mbox{\scriptsize dust}}=$20 K, \amax$=$0.1 mm].
 The walkers were iterated with 500 steps assuming flat priors; in the end, the results from the first 100 steps were discarded.
These fittings achieved good convergence in the offset range of [-10, 10] AU except at the central location (i.e., offset$=$0 AU).
Figure \ref{fig:corner} shows the corner plot produced from the MCMC fittings at the 4.1 AU offset as an example of the convergence.
The derived profiles of $T_{\mbox{\scriptsize dust}}$ and \amax from the MCMC fittings are presented in the top and bottom panels of Figure \ref{fig:profiles}; the inferred $\alpha$ values from the MCMC fittings are presented in the middle panel of Figure \ref{fig:profiles} but only for the offset range where the fittings converged well.
Following \citet{Tsukagoshi2016}, the top panel of Figure \ref{fig:profiles} also presents the $T_{\mbox{\scriptsize dust}}(R)$=22 [K]$\times$(R/10 [AU])$^{-0.4}$ and 28 [K]$\times$(R/10 [AU])$^{-0.4}$ midplane dust temperature profile models suggested from \citet{Andrews2012ApJ,Andrews2016ApJ}, where $R$ denotes the radius.
Note that the evaluation of these $T_{\mbox{\scriptsize dust}}(R)$ models did not consider dust scattering with the potentially radially varying \amax.

Results from the MCMC fittings show radially decreasing $T_{\mbox{\scriptsize dust}}$, which is everywhere higher than the observed $T_{\mbox{\scriptsize br}}$ at Bands 4 and 6 but yet appear reasonable.
Values of the derived \amax radially decrease from $\sim$100 $\mu$m to $\sim$20 $\mu$m.

The dominant errors of the derived $T_{\mbox{\scriptsize dust}}$ and \amax are systematic, which were induced by the uncertainties of $\kappa_{\nu}^{\mbox{\scriptsize abs}}$ and $\kappa_{\nu}^{\mbox{\scriptsize sca,eff}}$.
They depend on the dust composition and the exact form of grain size distribution \citep[e.g.][]{Sierra2017,Soon2017}, which are beyond the scope of the present work and are not quantitatively assessed.
In addition, the MCMC fittings have poor convergence outside of the offset range of [-10, 10] AU and at the central location. 
These poor convergences can be understood, since outside of the offset range of [-10, 10] AU, the observed $\alpha$ is becoming higher than 2.0, and that the optically thick assumption may not be valid at Band 4 (also see Figure 3 of \citealt{Tsukagoshi2016}).
In addition, TW\,Hya presents a low density cavity around the central location (Figure \ref{fig:alma}; see also \citealt{Andrews2016ApJ}).
The measured $T_{\mbox{\scriptsize br}}$ at the central location at 0$''$.085 resolution was, therefore, subject to significant beam dilution.
This led to degenerate fitting results of MCMC, which nevertheless reflect that the actual $T_{\mbox{\scriptsize dust}}$ should be higher than the beam diluted $T_{\mbox{\scriptsize br}}$ measurements.

Why can the application of the geometrically thin dust slab solution of \citet{Birnstiel2018} be self-consistently a good approximation?
Does scattering of the warm dust emission from the central part of the disk in turn steepen the spectral index?
We argue that the geometrically thin dust slab solution is indeed a good approximation for the case of TW\,Hya since the derived temperature variations in the region of our interests (e.g., $\sim$0-10 AU radii) is not large.
This is partly because TW\,Hya does not have a hot inner disk which is luminous at (sub)millimeter bands. 
Instead, the (sub)millimeter images of TW\,Hya present an inner cavity.
When the observed temperature variations and temperature gradients are not huge, and when the disk is geometrically thin, it is possible break down the global radiative transfer solution to a quasi-local problem.
In the case of a small temperature gradient,
we can consider the temperature of the thin slab to be {\it locally uniform}.
The adjacent disk components which are emitting at very different temperatures would have rather large spatial separations from the local component of interest, will see the local component at an asymptotically small solid angle, and hence cannot contribute to significant scattered light flux.

To verify these arguments, we have carried out simple three dimensional radiative transfer simulations using the RADMC-3D code\footnote{http://www.ita.uni-heidelberg.de/~dullemond/software/radmc-3d/}, and compared the results from simulations with and without switching on dust scattering.
In our RADMC-3D models, the radial gas column density ($\Sigma_{g}$) profile was assumed to be
\begin{equation}
  \Sigma_{g} \mbox{ [g\,cm$^{-2}$]}  = 4\cdot10^{3} \times (\frac{r}{\mbox{[AU]}})^{-0.5},
\end{equation}
where $r$ is the projected radius on the disk midplane.
The gas volume density ($\rho$) was estimated based on
\begin{equation}
    \rho \mbox{[g\,cm$^{-3}$]} = \Sigma_{g} \cdot \frac{1}{\sqrt{2\pi h}}e^{-\frac{z^2}{2h}},
\end{equation}
where $z$ is the vertical offset from the disk midplane, and $h$ is the characteristic disk scale height which we assumed as
\begin{equation}
    h \mbox{[AU]} = 0.05\cdot(\frac{r}{\mbox{[AU]}})^{1.1}.
\end{equation}
We truncated the column density profile interior to the 1 AU radius to mimic the presence of a inner cavity in TW\,Hya; our simulation covered a up to 20 AU radius.
We assumed a constant 0.01 dust-to-gas mass ratio, and assumed a constant \amax$=$0.1 mm.
Our dust density model is therefore a geometrically thin disk with modestly small flaring, which is very optically thick in the inner 1$\sim$10 AU region and becomes optically thinner at outer radii.
Examining the geometrically thin assumption requires intensive simulations of dust grain growth and dust vertical settling, which is by itself a developing research field and is well beyond the scope of the present paper.
We assumed the dust temperature to be
\begin{equation}
    T_{\mbox{\scriptsize dust}}(r, \phi, z)= 50 \mbox{ [K]} \times ( \frac{r}{\mbox{[10 AU]}} )^{-0.4},
\end{equation}
where $\phi$ is the azimuthal angle.
Our simulations assumed similar temperature gradients to what was actually observed from TW\,Hya but a higher absolute temperature scale.
This was because our main purpose is to test whether or not including dust scattering can indeed lead to the anomalously low $\alpha$ values.
Using higher dust temperatures can avoid producing low $\alpha$ values due to non-Rayleigh-Jeans effects.
In addition, instead of evaluating dust temperature based on radiative transfer, we used the assumed radial temperature profile.
Physically, this was because on the spatial scales of our interests, how dust can be heated due to viscous dissipation is not yet certain.
In addition, to simulate anisotropic dust scattering in the optically very thick limit, we need to use full three-dimensional grids with rather small grid sizes, which makes the precise temperature evaluation computationally expensive and unfeasible for us.
On the other hand, we do not want the simulations with and without dust scattering to converge to different temperature profiles, which will in turn confuse the discussion about the effects of dust scattering on $\alpha$.

\begin{figure}
   \vspace{-1.3cm}
   \hspace{-1.5cm}
    \begin{tabular}{ p{10cm} }
        \includegraphics[width=10.5cm]{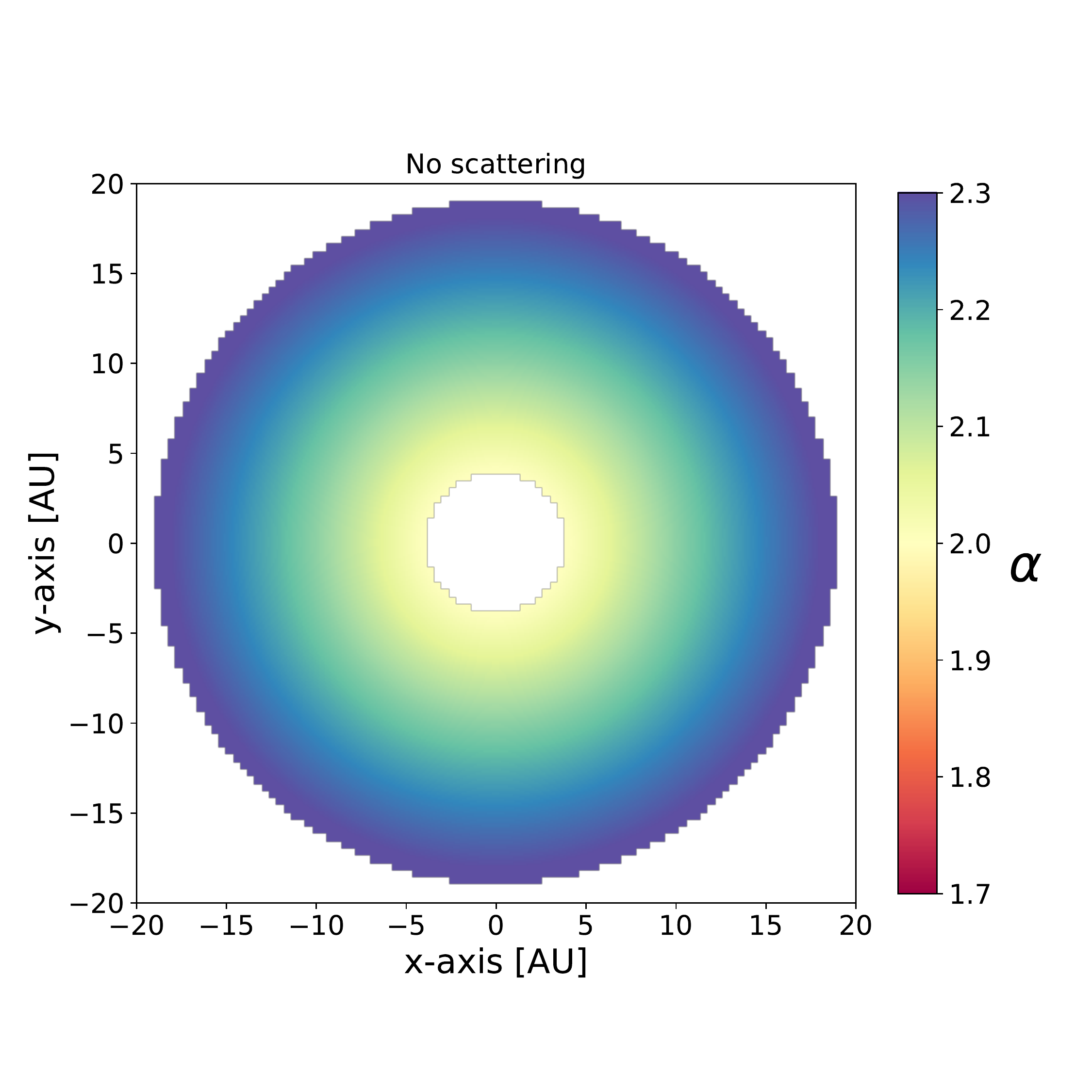} \\  
        \vspace{-2.5cm}\includegraphics[width=10.5cm]{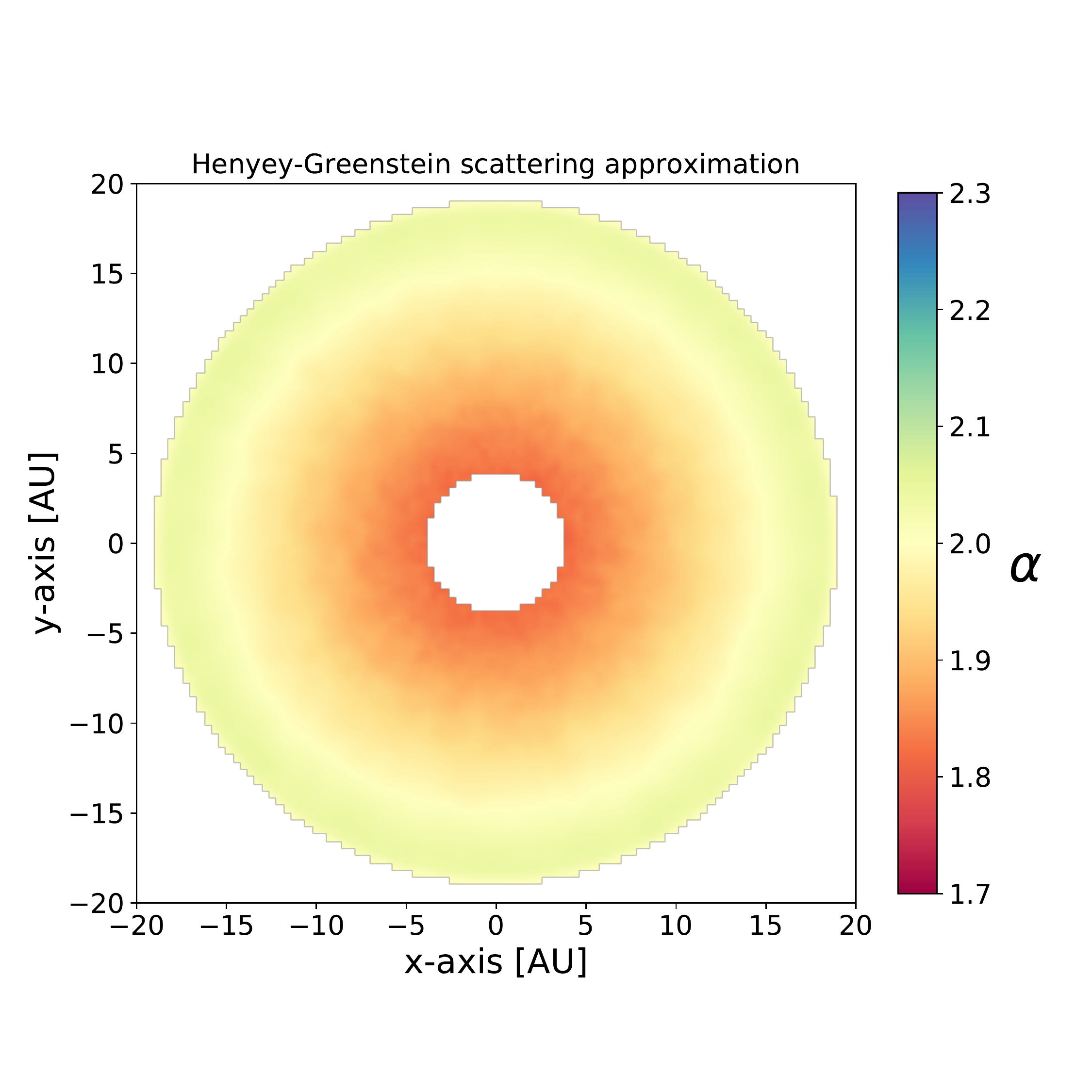} \\ 
    \end{tabular}
    \vspace{-1cm}
    \caption{The spectral index ($\alpha$) in between 144.988 GHz and 232.990 GHz derived from the RADMC-3D simlations for a face-on disk. Top and bottom panels show the cases with no scattering and with anisotropic scattering evaluted based on Henyey-Greenstein approximation, respectively. These simulations were based on the identical assumption of density and temperature distributions (see Section \ref{sec:analysis}).}
    \label{fig:radmc3d}
\end{figure}

The spatial grids of our simulations were defined in spherical coordinates with uniform intervals of polar angle, azimuthal angle, and logged radius.
Using the RADMC-3D code, we derived the pole-on view of the disk at 232.990 GHz and 144.988 GHz.
The simulated images have some numerical errors inward of the $\sim$3 AU radius due to the very rapid changes of dust volume density with the radius and polar angle, which can lead to $\sim\pm$0.02 errors of the derived spectral indices.
Therefore, we masked the inner 3.2 AU radius in the simulated images.
Figure \ref{fig:radmc3d} shows the derived $\alpha$ distributions from these simulations in the cases with and without switching on scattering.
Indeed, in the case without switching on scattering, $\alpha$ converges to 2.0 in the innermost, high optical depth region; when scattering was switched on, $\alpha$ can converge to values lower than 2.0, which supports our arguments about the appropriateness of applying the analytic thin slab solution.

\section{Discussion}\label{sec:discussion}

\citet{Tsukagoshi2016} reported that $\alpha$ has a value $\sim$3.7 at the $\sim$22 AU gap where the dust emission is relatively optically thin.
This result is consistent with \amax$<$0.1 mm \citep[see Figure 4 of][]{Birnstiel2018}, and can be reconciled with the \amax derived by the present work at smaller radii without requiring a rapid spatial variation of \amax.
\citet{Tsukagoshi2016} suggested a shortage of millimeter size grains in the 22 AU gap.
With the present work, it is also not clear where the millimeter size grains are presented inwards of the 22 AU gap.
Physically, even in the case that grown dust can efficiently form in regions inwards of the 22 AU gap, whether or not we can detect the dust that has grown to these with the presented observations remain questionable.
For example, the simulations of \citet{Vorobyov2018A&A} have shown that under certain physical conditions, dust that has grown to larger sizes can have rapid radial migration and can be trapped in regions which have too-small areas to be probed by observations.
The observations may also preferentially detect small dust grains at the scattering surface, due to the vertical settling of large dust grains \citep[e.g.,][]{Yang2017MNRAS,Hull2018,Dent2019}.

The \amax values derived in the present work (Figure \ref{fig:profiles}) have no tension with those derived from the previous (sub)millimeter polarimetric observations \citep[50-150 $\mu$m;][]{Kataoka2016ApJa,Kataoka2016ApJb,Hull2018}.
In this sense, the presented $\alpha$ values and dust polarization in HD\,163296 \citep{Dent2019} may be independent indicators of 10-100 $\mu$m maximum grain sizes.
In fact, resolving $\alpha$ at multiple wavelengths may serve as a cheap (in terms of observing time) auxiliary method to help assess whether or not the observed dust linear polarization at a specific wavelength is dominated by dust scattering.

On the other hand, assuming that \amax is still smaller in Class 0/I YSOs than in Class II protoplanetary disks, the previously observed dust linear polarization from Class 0/I YSOs may be preferably explained by aligned dust grains, which was supported by the highly consistent polarization percentages and position angles over broad ranges of wavelengths \citep[e.g.,][]{Liu2016ApJ,Alves2018A&A,Liu2018A&A2,Sadavoy2018ApJ}.
The hypothesis of small \amax values was also independently supported by astrochemical studies \citep{Harada2017ApJ}.

Finally, we note that when albedo is high, the observed dust brightness temperature can be considerably lower than the actual (or expected) dust temperature even when the dust optical depth is much higher than 1 \citep{Birnstiel2018}.
When fitting the millimeter SED with a program that does not take scattering opacity into account, the fittings may be driven to conclude optically thin dust with significant grain growth, which can, in turn, lead to an underestimate of the total mass of solids.

\acknowledgments %
HBL is extremely grateful to Dr. Dominique Segura-Cox for her helps with English editing.
HBL thanks the referee for the comments made from very critical thinking.
This paper makes use of the following ALMA data: ADS/JAO.ALMA \#2013.1.00114.S,  \#2015.1.00005.S. ALMA is a partnership of ESO (representing its member states), NSF (USA) and NINS (Japan), together with NRC (Canada), MOST and ASIAA (Taiwan), and KASI (Republic of Korea), in cooperation with the Republic of Chile. The Joint ALMA Observatory is operated by ESO, AUI/NRAO and NAOJ.
This work has made use of data from the European Space Agency (ESA) mission
{\it Gaia} (\url{https://www.cosmos.esa.int/gaia}), processed by the {\it Gaia}
Data Processing and Analysis Consortium (DPAC,
\url{https://www.cosmos.esa.int/web/gaia/dpac/consortium}). Funding for the DPAC
has been provided by national institutions, in particular the institutions
participating in the {\it Gaia} Multilateral Agreement.
HBL thanks Yasuhiro Hasegawa, Ryo Tazaki, and Mario Flock for the very useful discussion.
H.B.L. is supported by the Ministry of Science and
Technology (MoST) of Taiwan (Grant Nos. 108-2112-M-001-002-MY3 and 108-2923-M-001-006-MY3).

\facility{ALMA}
\software{CASA \citep{McMullin2007}, Numpy \citep{VanDerWalt2011}, emcee \citep{Foreman-Mackey2013PASP}, RADMC-3D (Dullemond et al. in prep.) }


\appendix

\section{Reproducing ALMA measurements}\label{appendix}

\begin{figure*}[h]
\hspace{-2.9cm}
  \begin{tabular}{ p{9cm} p{9cm} }
    \includegraphics[width=11cm]{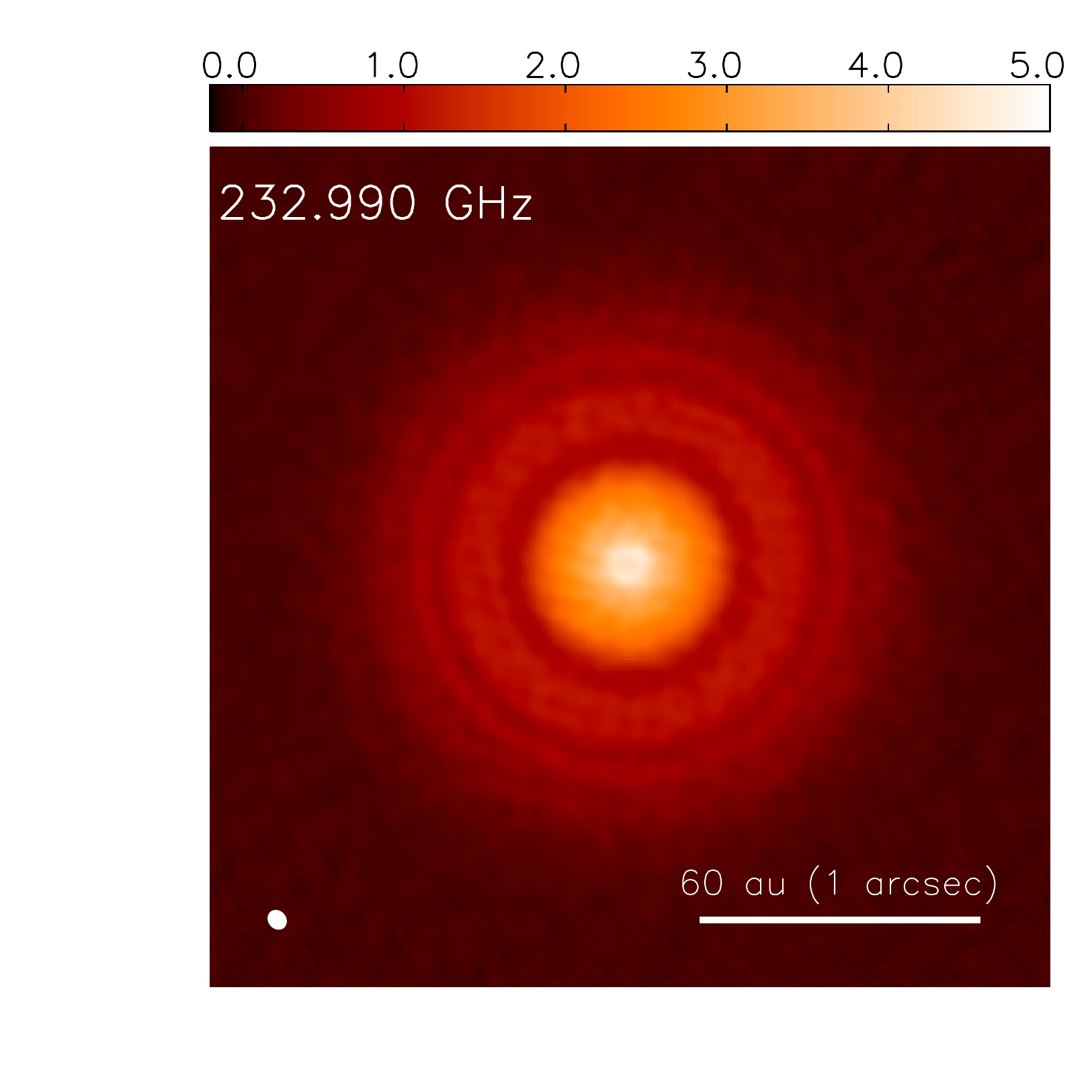} &
    \includegraphics[width=11cm]{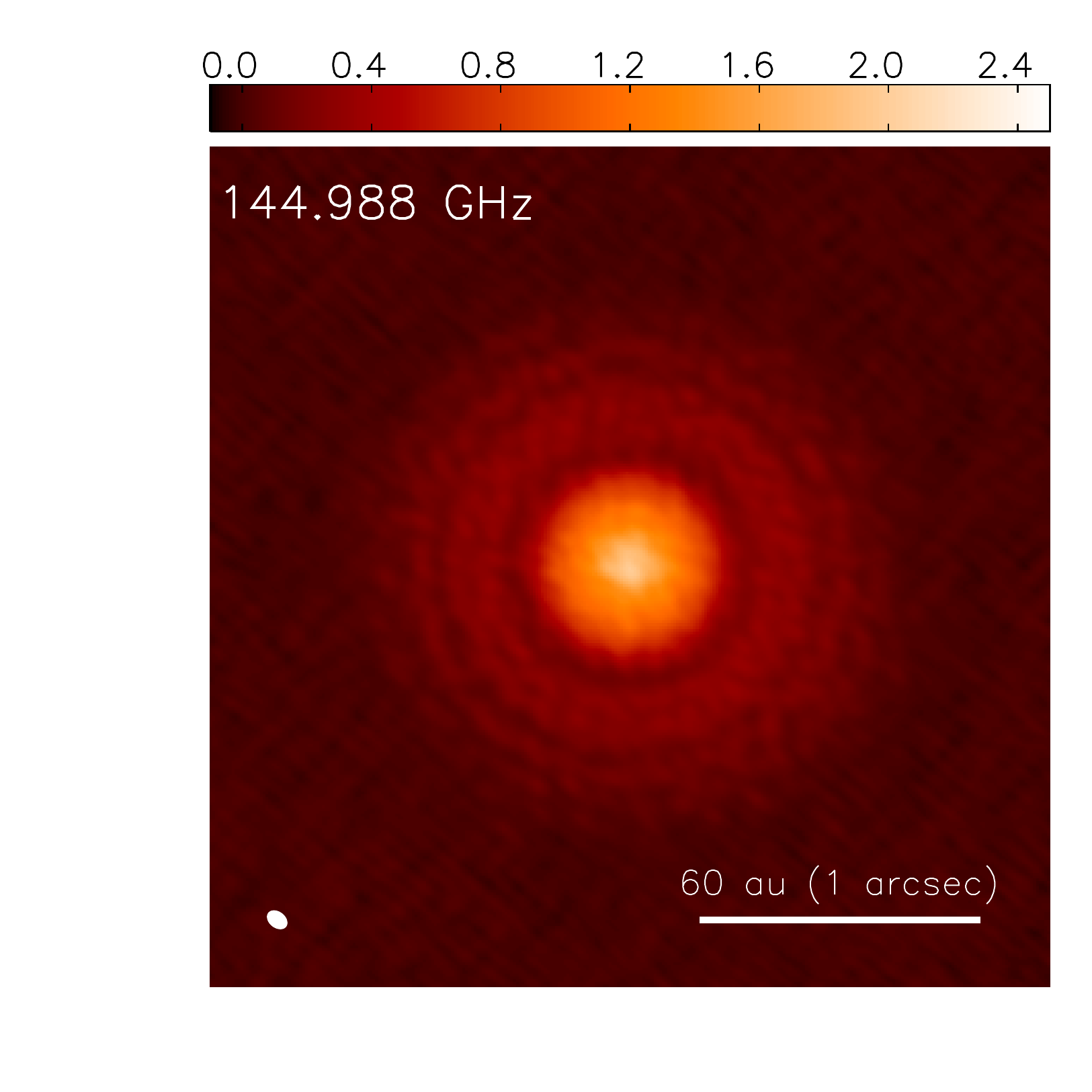} \\
  \end{tabular}
\vspace{-0.8cm}
\caption{
  ALMA 232.990 GHz (Band 6) and 144.988 GHz (Band 4) images of TW\,Hya in units of mJy\,beam$^{-1}$.
  The synthesized beams at Band 6 and 4 are \beam$=$0$''$.075$\times$0$''$.064 (P.A.=39$^{\circ}$) and \beam$=$0$''$.081$\times$0$''$.058 (P.A.=52$^{\circ}$), respectively, and are shown in bottom left of each panel.
  The RMS noise level of the Band 6 and 4 images are 23 $\mu$Jy\,beam$^{-1}$ and 14 $\mu$Jy\,beam$^{-1}$, respectively.
        }
   \label{fig:alma}
\end{figure*}

The archival ALMA Band 4 and 6 data were re-calibrated and phase self-calibrated following the strategy outlined in Section 2 of \citet{Tsukagoshi2016}, using the CASA software package v5.4.0  \citep{McMullin2007}.
The continuum data were derived using the CASA-{\tt uvcontsub} task.
The Band 4 and Band 6 continuum data were imaged separately, using the multi-frequency synthesis (MFS) method.
Unlike \citet{Tsukagoshi2016}, this work employed {\it nterm}=1 in MFS and did not employ multiscale clean, to avoid the systematic flux errors induced by spectral index errors and by non-local imaging artifacts.
The Band 6 image achieved a \beam$=$0$''$.075$\times$0$''$.064 synthesized beam and a 23 $\mu$Jy\,beam$^{-1}$ root-mean-square (RMS) noise level; the Band 4 image achieved a \beam$=$0$''$.081$\times$0$''$.058 synthesized beam and a 14  $\mu$Jy\,beam$^{-1}$ RMS noise level.
The final images achieved are presented in Figure \ref{fig:alma}. 
Afterward, these images were smoothed to have 0$''$.085 ($\sim$5.1 AU) circular beams before the analysis in this work.

\end{document}